\documentclass[aps,groupedaddress,showpacs,floatfix]{revtex4}
\usepackage{amssymb}
\usepackage{amsmath}
\usepackage{graphicx}
\begin{document}

\title{Two-time distribution functions in the Gaussian model
of randomly forced Burgers turbulence}

\author{Victor Dotsenko}

\affiliation{LPTMC, Universit\'e Paris VI, 75252 Paris, France}

\affiliation{L.D.\ Landau Institute for Theoretical Physics,
 119334 Moscow, Russia}

\date{\today}

\begin{abstract}

The problem of randomly forced Burgers turbulence is 
considered in terms of the toy Gaussian Larkin model of directed polymers. 
In terms of the replica technique
the explicit expressions for the two-time four-point free energy
distribution function is obtained which makes possible to derive
the exact result for the two-time velocity distribution function
in the corresponding Burgulence problem.

\end{abstract}


\maketitle

\section{Introduction}

The  KPZ problem \cite{KPZ} describing the growth in time of an interface
in the presence of noise
have been the subject of intense investigations during the past almost three
decades 
\cite{hh_zhang_95,burgers_74,kardar_book,hhf_85,numer1,numer2,kardar_87,bouchaud-orland,Brunet-Derrida,
Johansson,Prahofer-Spohn,Ferrari-Spohn1}.
During last five years the major progress in solving this problem has been achieved which resulted
in derivation of the exact solutions for various types of the free energy
probability distribution function (PDF)
\cite{KPZ-TW1a,KPZ-TW1b,KPZ-TW1c,KPZ-TW2,BA-TW1,BA-TW2,BA-TW3,LeDoussal1,LeDoussal2,end-point,goe,LeDoussal3,Corwin,Borodin,
Prolhac-Spohn,2pointPDF,Imamura-Sasamoto-Spohn,Prolhac-Spohn-N,2time,Npoint}.

In this paper I would like to consider the possibility to apply the ideas and technique developed in the studies of the 
KPZ problem for formally equivalent problem of the randomly forced Burgers turbulence (the "Burgulence" problem).
Here one considers a velocity field $v(x,t)$ governed by the Burgers equation
\begin{equation}
   \label{1}
\partial_{t} v(x,t) + v(x,t)\partial_{x} v(x,t) = 
\nu  \partial^{2}_{x} v(x,t) +  \zeta(x,t)
\end{equation}
where the parameter $\nu$ is the viscosity and $\zeta(x,t)$ is the Gaussian distributed random force 
which is $\delta$-correlated in time and which is characterized by finite correlation length $\xi$ in space: 
$\overline{\zeta(x,t) \zeta(x',t')} = u \delta(t-t') U[(x-x')/\xi]$. Here U(x) is a smooth function decaying to zero fast
enough at large arguments and the parameter $u$ is the injected energy density. In this problem one 
would like to derive e.g. the the probability distribution functions of the velocity gradients 
$P[\partial_{x} v(x,t)]$ or two-points distribution function $P[v(x,t), v(x',t)]$ at scales smaller
than the length scale $\xi$ of the stirring force $\zeta$
(see e.g. \cite{burgers_74,Sinai,Bouch-Mez-Par,burgulence} and references there in).

Formally the above problem is equivalent to the KPZ equation as well as to the (1+1) directed
polymers. Indeed, redefining $v(x,t) = -\partial_{x} F(x,t)$
and $\zeta(x,t) = -\partial_{x} V(x,t)$ one gets the KPZ equation for the interface profile
$F(x,t)$ (which is the free energy of (1+1) directed polymers):
\begin{equation}
   \label{2}
\partial_{t} F(x,t) = \frac{1}{2} \bigl(\partial_{x} F(x,t)\bigr)^{2} 
- \frac{1}{2} T \partial^{2}_{x} F(x,t) -  V(x,t)
\end{equation}
where $T = 2\nu$ is the temperature parameter of the directed polymer problem and $V(x,t)$ is the Gaussian
distributed random potential.

The idea of a new approach to the Burgulence problem which I would like to demonstrate in this paper
is in the following. According to the above definitions the velocity field can be
represented as 
\begin{equation}
\label{3}
v(x,t)=\lim_{\epsilon\to 0} [F(x+\epsilon,t) - F(x,t)]/\epsilon
 \end{equation}
Thus, deriving the four-point KPZ probability distribution function
${\cal P}[F(x+\epsilon,t), F(x,t), F(x'+\epsilon,t'),F(x',t')]$ and taking the limit $\epsilon \to 0$
one could hopefully obtain the result for $P[v(x,t),v(x',t')]$.
The only "little problem" is that unlike the usual KPZ studies operating with the $\delta$-correlated in space
random potential, in the Burgulence problem one is mainly interested in the spatial scales comparable
or much smaller than the random potential correlation length $\xi$. In other words, in this  
approach, first one has to study KPZ problem with random potentials having {\it finite} correlation length.
In the present study (as a matter of "warming up" exercise) I'm going to consider another "extreme case" 
in which the random potential $V(x,t)$ of the KPZ problem is changed by it's linear approximation: 
$V(x,t) \to \zeta(t) x$ where $\zeta(t)$ is Gaussian distributed random force. In this case we obtain 
the model introduced by Larkin \cite{Larkin, Larkin-Ovchinnikov} long time ago
to study small scale displacements of directed polymers. In this approximation the model becomes Gaussian
and therefore exactly solvable. Nevertheless, the statistical properties of its free energy (as well as some others
quantities)  turn out to be rather non-trivial (see e.g \cite{Gorochov-Blatter,gaussian,replicas}).
For that reason this model hopefully could serve as a good ground for testing various approaches developed 
in the recent KPZ studies.

In  Section II we introduce the model and formulate the main ideas of the replica approach which will be used 
in the further calculations of the two-time free energy distribution functions (for the directed polymer model)
and the corresponding velocity distribution function (in the Burgulence problem).
In Section III as the matter of the demonstration of the replica technique the explicit expression 
for the two-time (two-point) free energy distribution function is derived.
In Section IV the two-time four-point free energy distribution function is calculated and the 
corresponding two-time velocity distribution function $P[v(x,t),v(x',t')]$ of the Burgulence problem
is obtained. Further perspectives of the present approach is discussed in Section V.



\section{The model and the method}

In this paper we consider the model of one-dimensional directed polymers 
defined in terms of an elastic string $\phi(\tau)$
directed along the $\tau$-axes within an interval $[0,t]$ which passes through a random medium
described by a random potential $V(\phi,\tau) \; = \; \zeta(\tau) \phi$. The energy of a given polymer's trajectory
$\phi(\tau)$ is given by the Hamiltonian
\begin{equation}
   \label{4}
   H[\phi(\tau), V] = \int_{0}^{t} d\tau
   \Bigl\{\frac{1}{2} \bigl[\partial_\tau \phi(\tau)\bigr]^2
   + \zeta(\tau) \phi(\tau) \Bigr\};
\end{equation}
where the random force $\zeta(\tau)$ is described by the Gaussian distribution 
with a zero mean $\overline{\zeta(\tau)}=0$ 
and the $\delta$-correlations: 
\begin{equation}
   \label{5}
{\overline{\zeta(\tau)\zeta(\tau')}} = u \delta(\tau-\tau')
\end{equation}
The parameter $u$ describes the strength of the disorder.

For the fixed boundary conditions, $\phi(0) = y; \; \phi(t) = x$, the partition function
of the model (\ref{4}) is
\begin{equation}
\label{6}
   Z(x|y; t) = \int_{\phi(0)=y}^{\phi(t)=x}
              {\cal D} \phi(\tau)  \;  \mbox{\Large e}^{-\beta H[\phi]}
\; = \; \exp\bigl[-\beta F(x|y; t)\bigr]
\end{equation}
where $\beta$ is the inverse temperature and $F(x|y; t)$ is the free energy.
In the replica approach one consider the average of the $N$-th power of the 
above partition function:
\begin{equation}
 \label{7}
 \overline{Z^{N} (x|y; t)} \; \equiv \; Z(N; x|y ; t) \; = \; \overline{\exp\bigl[-\beta N F(x|y; t)\bigr]}
\end{equation}
where $\overline{(...)}$ denotes the average over the random force $\zeta$.
Simple Gaussian averaging yields:
\begin{equation}
   \label{8}
Z(N; x|y ; t)  \; = \; \prod_{a=1}^{N} \Biggl[
\int_{\phi_{a}(0)=y}^{\phi_{a}(t)=x} {\cal D} \phi_{a}(\tau) \Biggr] \;
   \exp\Bigl[-\beta H_{n}[{\boldsymbol \phi}]  \Bigr]
\end{equation}
where
\begin{equation}
 \label{9}
    H_{N}[{\boldsymbol \phi}] \; = \;
   \frac{1}{2} \int_{0}^{t} d\tau \Biggl(
   \sum_{a=1}^{N} \bigl[\partial_\tau \phi_{a}(\tau)\bigr]^2 
   - \beta u \sum_{a, b}^{N} \phi_{a}(\tau) \phi_{b}(\tau) \Biggr) 
\end{equation}
is the Gaussian replica Hamiltonian.

Introducing the free energy distribution function, $P_{x|y; t}(F)$, the relation (\ref{7})
can be represented as follows:
\begin{equation}
 \label{10}
 Z(N; x|y ; t) \; = \; \int_{-\infty}^{+\infty} dF \; P_{x|y; t}(F) \exp\bigl[-\beta N F\bigr]
\end{equation}
which is the Laplace transform of the distribution function, $P_{x|y; t}(F)$ with respect to 
the parameter $\beta N$. 
In the lucky case when the moments of the partition function $Z(N; x|y ; t)$ allows an analytic continuation
from integer to arbitrary complex values of the replica parameter $N$ the above relation makes possible
to reconstruct the probability distribution function $P_{x|y; t}(F)$ via the inverse Laplace transform:
\begin{equation}
 \label{11}
 P_{x|y; t}(F) \; = \; \int_{-i\infty}^{+i\infty} \frac{ds}{2\pi i} \; Z\bigl(\frac{s}{\beta}; x|y ; t\bigr) \; 
                       \exp\bigl(s F\bigr)
\end{equation}
In the present model this distribution can be computed explicitly and the resulting
function $ P_{x|y; t}(F)$ turns out to be rather non-trivial \cite{Gorochov-Blatter,gaussian,replicas}.

Besides the replica partition function eq.(\ref{8}) 
it is convenient to introduce the $N$-particle "wave function":
\begin{equation}
 \label{12}
 \Psi_{N}\bigl[{\bf x}|{\bf y}; t\bigr] \; = \; \prod_{a=1}^{N} 
 \Biggl[
\int_{\phi_{a}(0)=y_{a}}^{\phi_{a}(t)=x_{a}} 
   {\cal D} \phi_{a}(x)
   \Biggr] \;
   \exp\Bigl[-\beta H_{N}[{\boldsymbol \phi}]  \Bigr]
\end{equation}
where the Hamiltonian $H_{N}[{\boldsymbol \phi}]$ is given in eq.(\ref{9}).
In the present model the function $\Psi_{N}\bigl[{\bf x}|{\bf y}; t\bigr]$ can be computed
explicitly (see e.g. \cite{replicas}):
\begin{equation}
\label{13}
 \Psi_{N}\bigl[{\bf x}|{\bf y}; t\bigr] = C\bigl(N, t\bigr)  
 \exp\Biggl\{
 -\frac{\beta}{2t} \sum_{a=1}^{N} \bigl(x_{a} - y_{a}\bigr)^{2}  
 -\frac{\beta}{2t} A(N,t) \Biggl[\sum_{a=1}^{N} \bigl(x_{a} - y_{a}\bigr) \Biggr]^{2} 
 +\frac{\beta}{t} B(N,t) \sum_{a,b=1}^{N} x_{a} y_{b} 
\Biggr\}
\end{equation}
where
\begin{eqnarray}
 \label{14}
 C(N,t) &=& \Bigl(\frac{\beta}{2\pi t}\Bigr)^{N/2} \; 
            \sqrt{\frac{\sqrt{\beta N u t^{2}}}{\sin\bigl( \sqrt{\beta N u t^{2}}\bigr)}}
            \\
            \nonumber
            \\
 \label{15}
 A(N,t) &=& \frac{1}{N} \Biggl( 
            \frac{\sqrt{\beta N u t^{2}}}{\tan\bigl( \sqrt{\beta N u t^{2}}\bigr)} \; - \; 1
            \Biggr)
            \\
            \nonumber
            \\
 \label{16}
 B(N,t) &=& \frac{1}{N} \; \frac{\sqrt{\beta N u t^{2}}}{\sin\bigl( \sqrt{\beta N u t^{2}}\bigr)} \; 
                           \bigl[ 1 \; - \; \cos\bigl(\sqrt{\beta N u t^{2}}\bigr) \bigr]
\end{eqnarray}
Note that the above expression for the wave function eq.(\ref{13}) is valid
only at finite time interval: $t \; < \; \frac{\pi}{2} (\beta N u)^{-1/2} \equiv t_{c}(N) $. 
The reason is that due to specific form of the
interaction potentials in the replica Hamiltonian (\ref{9}) the directed polymers trajectories
go to infinity at {\it finite time} $t_{c}(N)$.  For the original physical system this phenomenon
is explained by the presence of the slowly decaying {\it left tail} \cite{Gorochov-Blatter,gaussian,replicas} 
of the free energy distribution function $ P_{x|y; t}(F)$ which
(according to the relation (\ref{10})) results in the divergence of all partition function moments   
$Z(N; x|y ; t)$ with $N > N_{c}(t) = \frac{\pi^{2}}{4} (\beta u t^{2})^{-1}$

\section{Two-time free energy distribution function}

For simplicity, in this section we consider the directed polymer problem with the zero boundary conditions:
$\phi(0) = \phi(t) = 0$. For a given realization of the random function $\zeta(\tau)$
let us denote by $F_{1}$ the free energy of the directed polymers with the (zero) ending point
at time $t$ and by $F_{2}$ the one of the directed polymers with the (zero) ending point
at time $t \, + \, \Delta t$.
According to the definition, eq.(\ref{6}), for the difference of these free energies, $F = F_{2} - F_{1}$
one has:
\begin{equation}
 \label{17}
 \exp\{-\beta F \} \; = \; Z^{-1}(0|0; \, t) \, Z(0|0; \, t+\Delta t) 
\end{equation}
Taking $n$-th power of the above relation and performing the averaging over quenched randomness we gets
\begin{equation}
 \label{18}
 \overline{\exp\{-\beta n F \}} \; = \; \overline{Z^{-n}(0|0; \, t) \, Z^{n}(0|0; \, t+\Delta t) }
\end{equation}
Introducing the two-time free energy difference probability distribution function
$P_{t,\Delta t}(F)$ in terms of the replica approach the above relation can be represented as 
follows:
\begin{equation}
 \label{19}
 \int_{-\infty}^{+\infty} dF \, P_{t,\Delta t}(F) \, \exp\{-\beta n F \} \; = \; 
 \lim_{N\to 0} \, \overline{Z^{N-n}(0|0; \, t) \, Z^{n}(0|0; \, t+\Delta t) }
\end{equation}
where according to the usual replica formalism, in the first step, the r.h.s of the above relation 
is computed for an arbitrary {\it integer} $N$ and then, at the second step, the obtained result is analytically
continued for arbitrary real $N$, and finally the limit $N\to 0$ is taken.

For an integer $N > n$ in terms of the wave function, eqs.(\ref{12})-(\ref{13}), the product of the
partition functions in the r.h.s. of eq.(\ref{19}) can be represented as follows:
\begin{equation}
 \label{20}
 \overline{Z^{N-n}(0|0;  t)  Z^{n}(0|0;  t+\Delta t) } \equiv
 Z(N, n; t, \Delta t)  =  
 \int_{-\infty}^{+\infty} dx_{1} ... dx_{n}  
 \Psi_{N}\bigl[\underbrace{0, ..., 0}_{N-n}, x_{1}, ..., x_{n}|{\bf 0}; t\bigr] \,
 \Psi^{*}_{n}\bigl[x_{1}, ..., x_{n}|{\bf 0}; \Delta t\bigr]
\end{equation}
where the second (conjugate) wave function represent the "backward" propagation
from the time moment $(t + \Delta t)$ to the previous time moment $t$.
Schematically the above expression is represented in Figure 1.

 \begin{figure}[h]
\begin{center}
   \includegraphics[width=12.0cm]{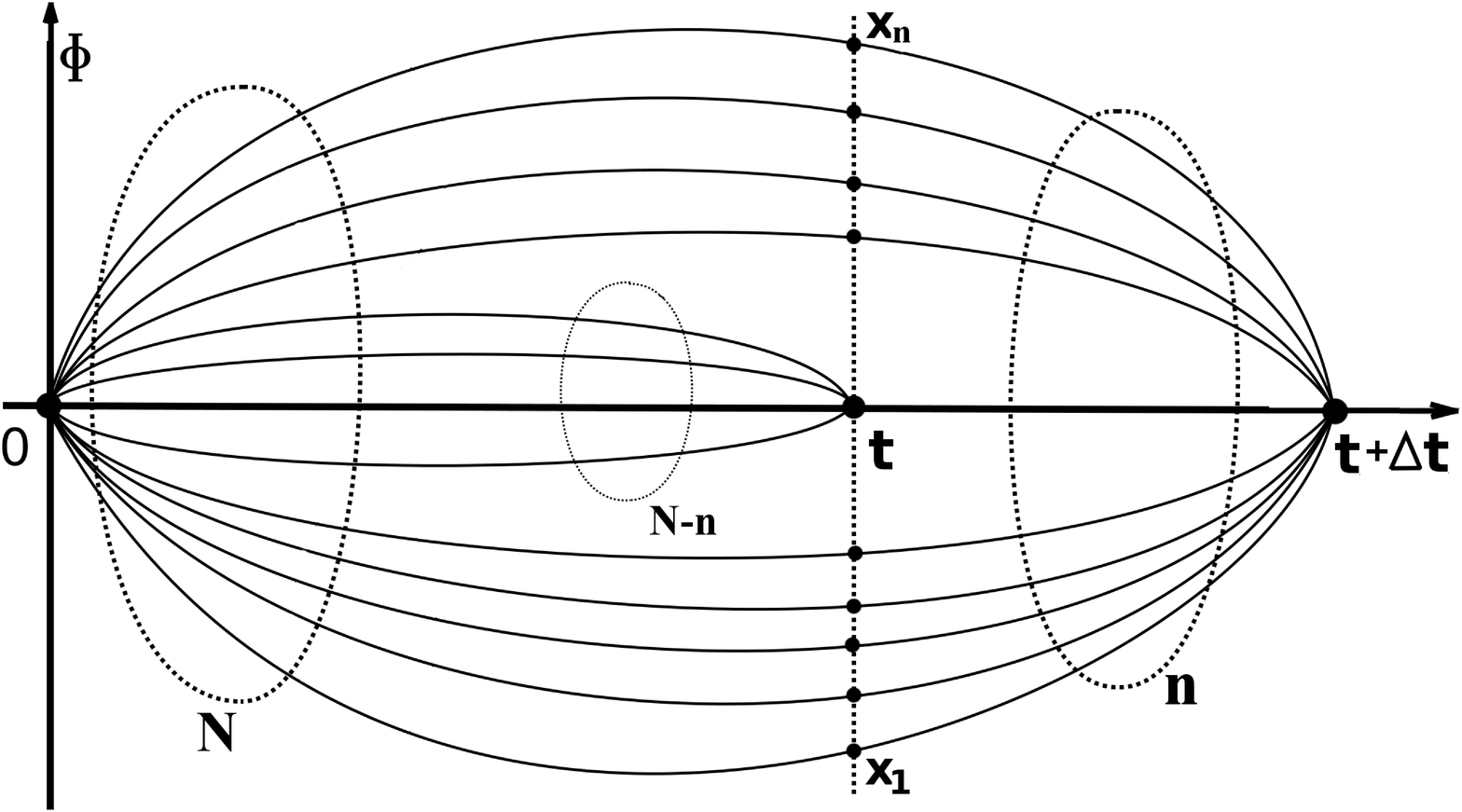}
\caption[]{Schematic representation of the directed polymer paths
corresponding to eq.(\ref{20})}
\end{center}
\label{figure1}
\end{figure}
 
Substituting eq.(\ref{13}) into eq.(\ref{20}) we get:
\begin{equation}
 \label{21}
 Z(N, n; t, \Delta t)  = 
 C(N,t) C(n,\Delta t) \int_{-\infty}^{+\infty} dx_{1} ... dx_{n} 
 \exp\Biggl\{
 -\frac{1}{2}\beta \sum_{a,b=1}^{n} 
 \Bigl[
 \frac{t + \Delta t}{t \Delta t} \delta_{ab} + 
 \frac{1}{t} A(N,t) +\frac{1}{\Delta t} A(n,\Delta t)
 \Bigr] x_{a} x_{b} 
 \Biggr\}
\end{equation}
Simple integration yields:
\begin{equation}
 \label{22}
 Z(N, n; t, \Delta t)  =  C(N,t) C(n,\Delta t)  
 \Bigl[\frac{2\pi t \Delta t}{\beta (t + \Delta t)}\Bigr]^{n/2} \; 
 \Biggl[
 1 \; + \; n\frac{t \Delta t}{t + \Delta t} \Bigl(\frac{1}{t} A(N,t) +\frac{1}{\Delta t} A(n,\Delta t)\Bigr)
 \Biggr]^{-1/2}
\end{equation}
Substituting here the explicit expressions (\ref{14}) and (\ref{15}), and taking the limit $N\to 0$ we obtain
\begin{equation}
 \label{23}
\lim_{N\to 0}  Z(N, n; t, \Delta t)  \; = \; 
\Bigl(\frac{t}{t + \Delta t}\Bigr)^{n/2}
\sqrt{\frac{
\sqrt{\beta n u (\Delta t)^{2}} \; (t + \Delta t)}{
\sin\bigl( \sqrt{\beta n u (\Delta t)^{2}}\bigr) 
\Bigl[
\Delta t - \frac{1}{3}\beta n u (\Delta t) t^{2} + 
t \frac{\sqrt{\beta n u (\Delta t)^{2}}}{\tan\bigl( \sqrt{\beta n u (\Delta t)^{2}}\bigr)}
\Bigr]}}
\end{equation}
substituting eq.(\ref{23}) into eq.(\ref{19}) and redefining:
\begin{eqnarray}
 \label{24}
 \Delta t \; = \; \xi \, t 
 \\
 \nonumber
 \\
 \label{25}
 \beta n u (\Delta t)^{2} \; = \; \omega
 \\
 \nonumber
 \\
 \label{26}
 F \; = \; u \xi^{2} t^{2} \, f
\end{eqnarray}
we get the following relation 
for the probability distribution function $P_{t,\xi} (f)$  
of the rescaled free energy $f$:
\begin{equation}
 \label{27}
 \int_{-\infty}^{+\infty} df \, P_{t,\xi}(f) \, \exp\{-\omega f \} \; = \; 
 \bigl(1 + \xi\bigr)^{-\frac{\omega}{2\beta u \xi^{2} t^{2}}}
 \frac{\omega^{1/4} \sqrt{1+ \xi}}{
 \sqrt{
\Bigl(\xi - \frac{\omega}{3\xi}\Bigr) \sin(\sqrt{\omega}) \; + \; \sqrt{\omega} \cos(\sqrt{\omega})}}
 \end{equation}
By inverse Laplace transform in the limit when both $t \to \infty$ and $\Delta t \to \infty$ (such that
the parameter ($\xi = \Delta t/t$ remains finite) we get the following universal 
result for the limiting two-time free energy distribution function:
\begin{equation}
\label{28}
\lim_{t\to\infty} P_{t,\xi}(f) \equiv {\cal P}_{\xi} (f) \; = \; 
\sqrt{1+\xi} \, 
\int_{-i\infty}^{+i\infty} \frac{d\omega}{2\pi i} \; 
\frac{\omega^{1/4} \exp\{\omega f\}}{
 \sqrt{
\Bigl(\xi - \frac{\omega}{3\xi}\Bigr) \sin(\sqrt{\omega}) \; + \; \sqrt{\omega} \cos(\sqrt{\omega})}}
 \end{equation}
It is interesting to note that this function (like its one-time  counterpart 
\cite{Gorochov-Blatter,gaussian,replicas})
is identically equal to zero at $f > 0$. 
Indeed, since at $f>0$ the function under the integral in the r.h.s of eq.(\ref{28}) quickly
goes to zero at $w \to -\infty$, the contour of integration in the complex plane 
can be safely shifted to $-\infty$, which means that $ {\cal P}_{\xi}(f > 0) \equiv 0$.

\section{Two-time velocity distribution function}

Velocity in the Burgers problem is given by the derivative of the free energy
of the directed polymer problem:
\begin{equation}
 \label{29}
 v(x,t) \; = \; -\frac{\partial F(x,t)}{\partial x} \; = \; 
             - \lim_{\epsilon \to 0} \frac{F(x+\epsilon) - F(x,t)}{\epsilon}
\end{equation}
Thus, to compute the two-point velocity distribution function in terms of the directed polymers, first, keeping 
$\epsilon$ finite we consider specially constructed four-point object (see below), and only in the final 
stage of calculations we take the limit $\epsilon \to 0$.

According to the relation (\ref{6})
\begin{equation}
 \label{30}
 \exp\{-\beta\bigl[F(x+\epsilon,t) - F(x,t)\bigr]\} \; = \; 
 \exp\{\beta\epsilon v(x,t)\} \; = \; 
 \frac{Z(x+\epsilon|0; \, t)}{Z(x|0; \, t)}
\end{equation}
Following the procedure described in the previous section we have:
\begin{eqnarray}
 \label{31}
 &&
\overline{
\exp\bigl\{ \beta n_{1} \epsilon v(x_{1},t) + \beta n_{2} \epsilon v(x_{2},t+\Delta t)\bigr\} } \; = \; 
\\
\nonumber
 \\
 \nonumber
&& \hspace{20mm} = \;
\lim_{N_{1}, N_{2} \to 0}
\overline{ 
Z^{n_{1}}(x_{1}+\epsilon|0; \, t) \, Z^{N_{1} - n_{1}}(x_{1}|0; \, t) 
Z^{n_{2}}(x_{2}+\epsilon|0; \, t + \Delta t) \, Z^{N_{2} - n_{2}}(x_{2}|0; \, t + \Delta t) }
\end{eqnarray}
Introducing two-time velocity distribution function $P_{x_{1},x_{2} t, \Delta t} (v_{1}, v_{2})$ 
the above relation can be represented as follows:
\begin{equation}
 \label{32}
 \int\int_{-\infty}^{+\infty} d v_{1} d v_{2} \; P_{x_{1},x_{2} t, \Delta t} (v_{1}, v_{2})
 \exp\bigl\{ \beta n_{1} \epsilon v_{1} + \beta n_{2} \epsilon v_{2}\bigr\} 
 \; = \;
 \lim_{N_{1}, N_{2} \to 0} \; {\cal Z}_{\epsilon}\bigl( N_{1}, n_{1}, N_{2}, n_{1}, x_{1}, x_{2}, t, \Delta t \bigr)
\end{equation}
where 
\begin{eqnarray}
 \label{33}
&& {\cal Z}_{\epsilon}\bigl( N_{1}, n_{1}, N_{2}, n_{2}, x_{1}, x_{2}, t, \Delta t \bigr) \; = 
 \\
\nonumber
 \\
 \nonumber
&& \hspace{20mm} = \; 
 \overline{ 
Z^{n_{1}}(x_{1}+\epsilon|0; \, t) \, Z^{N_{1} - n_{1}}(x_{1}|0; \, t) 
Z^{n_{2}}(x_{2}+\epsilon|0; \, t + \Delta t) \, Z^{N_{2} - n_{2}}(x_{2}|0; \, t + \Delta t) }
\end{eqnarray}
In terms of the wave function, eq.(\ref{12})-(\ref{13}),
\begin{eqnarray}
 \label{34}
&&
{\cal Z}_{\epsilon}\bigl( N_{1}, n_{1}, N_{2}, n_{2}, x_{1}, x_{2}, t, \Delta t \bigr) \; = 
 \\
\nonumber
 \\
 \nonumber
&&  = \; 
\int_{-\infty}^{+\infty} \, Dy \; 
\int_{-\infty}^{+\infty} \, Dz \; 
\Psi_{N_{1}+N_{2}}\bigl[\underbrace{x_{1}, ..., x_{1}}_{N_{1}-n_{1}}, 
                        \underbrace{x_{1}+\epsilon, ..., x_{1}+\epsilon}_{n_{1}},
                        y_{1}, ..., y_{N_{2}-n_{2}},
                        z_{1}, ..., z_{n_{2}} 
                        \, | \,{\bf 0}; 
                        \; t \bigr] \times
 \\
\nonumber
 \\
 \nonumber
&&  \hspace{35mm}  \times                       
 \Psi^{*}_{N_{2}}\bigl[y_{1}, ..., y_{N_{2}-n_{2}},
                       z_{1}, ..., z_{n_{2}} 
                       \,| \,
                       \underbrace{x_{2}, ..., x_{2}}_{N_{2}-n_{2}}, 
                       \underbrace{x_{2}+\epsilon, ..., x_{2}+\epsilon}_{n_{2}} ;
                       \, \Delta t\bigr]
\end{eqnarray}
where
\begin{eqnarray}
 \label{35}
 \int_{-\infty}^{+\infty} \, Dy \; &\equiv& \; \prod_{a=1}^{N_{2}-n_{2}} \int_{-\infty}^{+\infty} dy_{a}
 \\
\nonumber
 \\
 \nonumber
 \int_{-\infty}^{+\infty} \, Dz \; &\equiv& \; \prod_{a=1}^{n_{2}} \int_{-\infty}^{+\infty} dz_{a}
\end{eqnarray}
Schematically the expression in eq.(\ref{34}) is represented in Figure 2.

 \begin{figure}[h]
\begin{center}
   \includegraphics[width=13.0cm]{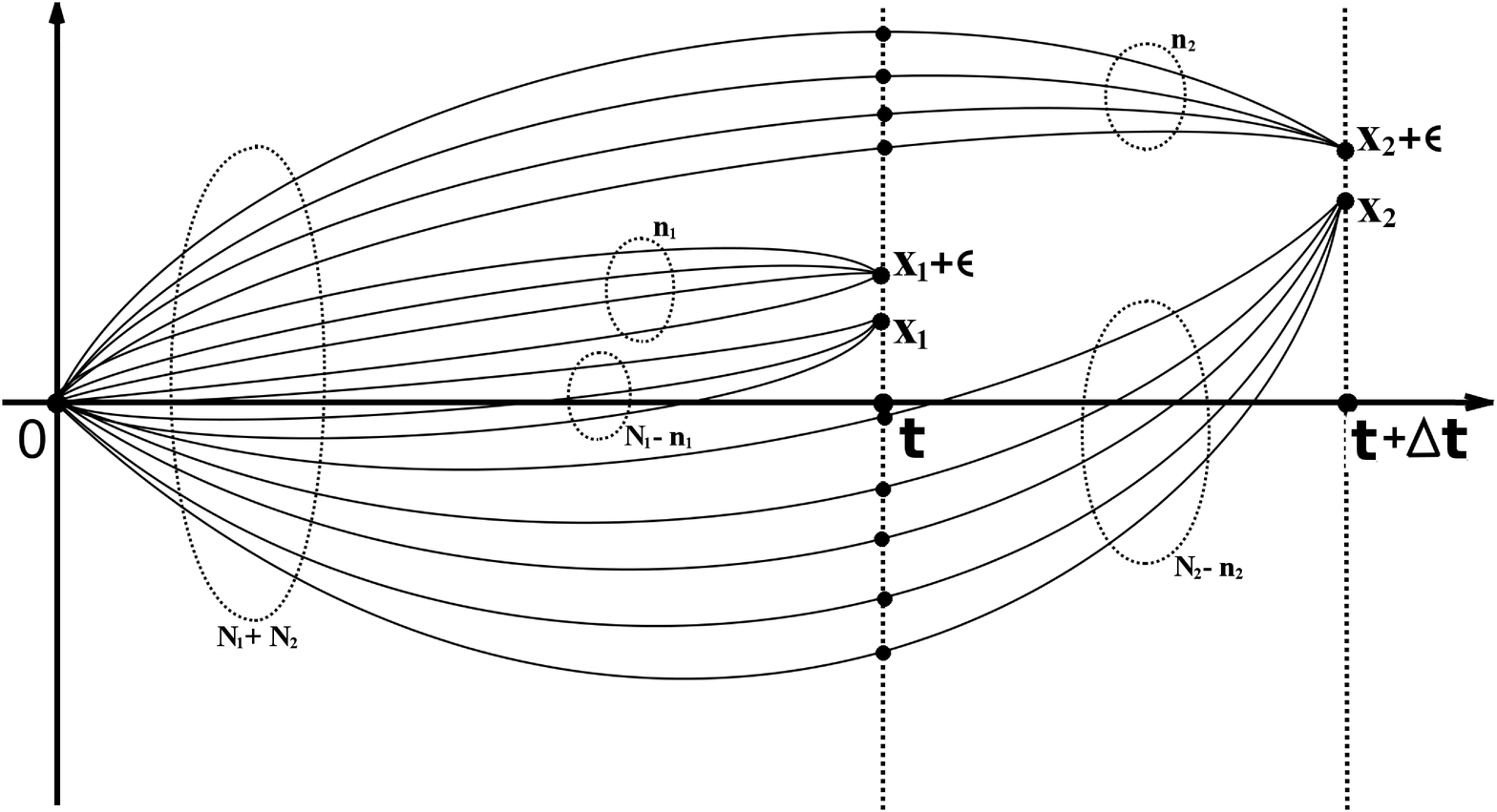}
\caption[]{Schematic representation of the directed polymer paths
corresponding to eq.(\ref{34})}
\end{center}
\label{figure2}
\end{figure}

Substituting the explicit expressions for the wave function (\ref{13}) into eq.(\ref{34}) 
 we get:
\begin{eqnarray}
\nonumber
&&
{\cal Z}_{\epsilon}\bigl( N_{1}, n_{1}, N_{2}, n_{2}, x_{1}, x_{2}, t, \Delta t \bigr) \; = \;
C(N_{1}+N_{2},t) \, C(N_{2}, \Delta t) \times
 \\
\nonumber
 \\
 \nonumber
&&  \times
\int_{-\infty}^{+\infty} \, Dy \; 
\int_{-\infty}^{+\infty} \, Dz \; 
\exp\Biggl\{
-\frac{\beta}{2t}
\Biggl[
(N_{1}-n_{1})x_{1}^{2} + n_{1} (x_{1}+\epsilon)^{2} + \sum_{a=1}^{N_{2}-n_{2}} y_{a}^{2} + \sum_{a=1}^{n_{2}} z_{a}^{2} 
\Biggr] \, -
 \\
\nonumber
 \\
 \nonumber
&& \hspace{23mm}
-\frac{\beta}{2t} A(N_{1}+N_{2}, t)
\Biggl[
(N_{1}-n_{1})x_{1} + n_{1} (x_{1}+\epsilon) + \sum_{a=1}^{N_{2}-n_{2}} y_{a} + \sum_{a=1}^{n_{2}} z_{a} 
\Biggr]^{2} \, -
 \\
\nonumber
 \\
 \nonumber
&& \hspace{23mm}
-\frac{\beta}{2\Delta t}
\Biggl[
\sum_{a=1}^{N_{2}-n_{2}} (y_{a}-x_{2})^{2} + \sum_{a=1}^{n_{2}} (z_{a}-x_{2}-\epsilon)^{2} 
\Biggr] \, -
 \\
\nonumber
 \\
 \nonumber
&& \hspace{23mm}
-\frac{\beta}{2\Delta t} A(N_{2}, \Delta t)
\Biggl[
\sum_{a=1}^{N_{2}-n_{2}} (y_{a}-x_{2}) + \sum_{a=1}^{n_{2}} (z_{a}-x_{2}-\epsilon) 
\Biggr]^{2} \, +
 \\
\nonumber
 \\
&& \hspace{23mm}
+ \frac{\beta}{\Delta t} B(N_{2}, \Delta t)
\Bigl(
\sum_{a=1}^{N_{2}-n_{2}} y_{a} + \sum_{a=1}^{n_{2}} z_{a}
\Bigr)
\Bigl[
(N_{2}-n_{2})x_{2} + n_{2} (x_{2}+\epsilon)
\Bigr]
\Biggr\}
\label{36}
\end{eqnarray}
where $A(N,t)$, $B(N,t)$ and $C(N,t)$ are given in eqs.(\ref{14})-(\ref{16}).
Introducing $N_{2}$-component vector ${\boldsymbol \chi} = \{y_{1}, ..., y_{N_{2}-n_{2}}, z_{1}, ..., z_{n_{2}}\}$
after simple algebra we get
\begin{eqnarray}
\nonumber
{\cal Z}_{\epsilon}\bigl( N_{1}, n_{1}, N_{2}, n_{2}, x_{1}, x_{2}, t, \Delta t \bigr) &=&
C(N_{1}+N_{2},t) \, C(N_{2}, \Delta t) 
\times
 \\
\nonumber
 \\
&\times&
\Biggl[\prod_{a=1}^{N_{2}} \int_{-\infty}^{+\infty} d\chi_{a}
\Biggr] \; 
\exp\Biggl\{
-\frac{1}{2} \sum_{a,b=1}^{N_{2}} T_{ab} \chi_{a} \chi_{b} \; + \; \sum_{a=1}^{N_{2}} L_{a} \chi_{a}
-\frac{1}{2} \beta  G
\Biggr\}
 \label{37}
\end{eqnarray} 
where
\begin{eqnarray}
\nonumber
 G &=&      \frac{1}{t} \bigl[ n_{1}(x_{1}+\epsilon)^{2} + (N_{1} - n_{1}) x_{1}^{2}\bigr] \; + \;
     \frac{1}{\Delta t} \bigl[ n_{2}(x_{2}+\epsilon)^{2} + (N_{2} - n_{2}) x_{2}^{2}\bigr] \; +
\\
\nonumber
 \\
&+& \frac{1}{t} A(N_{1} + N_{2},t) \bigl(N_{1} x_{1} + n_{1} \epsilon\bigr)^{2} \; + \;
    \frac{1}{\Delta t} A(N_{2},\Delta t) \bigl(N_{2} x_{2} + n_{2} \epsilon\bigr)^{2}  
\label{38}
\end{eqnarray}
and 
\begin{eqnarray}
 \label{39}
 T_{ab} &=& \gamma \, \delta_{ab} \; + \; \kappa
\\
\nonumber
\\ 
\label{40}
\gamma &=& \beta \frac{t + \Delta t}{t \Delta t}
\\
\nonumber
\\ 
\label{41}
\kappa &=& \frac{\beta}{t} A(N_{1}+N_{2}, t) \; + \; \frac{\beta}{\Delta t} A(N_{2}, \Delta t)
\\
\nonumber
\\ 
\label{42}
L_{a} &=& L \; + \; X_{a}
\\
\nonumber
\\ 
\label{43}
L &=& -\frac{\beta}{t} A(N_{1}+N_{2}, t) \bigl(N_{1} x_{1} + n_{1} \epsilon\bigr) \; + \; 
       \frac{\beta}{\Delta t} \bigl[ A(N_{2},\Delta t) + B(N_{2},\Delta t) \bigr] \bigl(N_{2} x_{2} + n_{2} \epsilon)
\\
\nonumber
\\ 
\label{44}
X_{a} &=& \left\{ \begin{array}{ll}
                 x_{2} + \epsilon,  & \mbox{for $a = 1, ..., n_{2}$}
 \\
                 x_{2},             & \mbox{for $a = n_{2}+1, ..., N_{2}$}
                            \end{array}
                            \right.       
\end{eqnarray}
Simple integration over $\chi$'s in eq.(\ref{37}) yields:
\begin{eqnarray}
\nonumber
{\cal Z}_{\epsilon}\bigl( N_{1}, n_{1}, N_{2}, n_{2}, x_{1}, x_{2}, t, \Delta t \bigr) \; &=& \;
C(N_{1}+N_{2},t) \, C(N_{2}, \Delta t) 
\times
 \\
\nonumber
 \\
&\times&
\exp\Biggl\{-\frac{1}{2} \beta  G
-\frac{1}{2} \mbox{Tr} \,\ln \hat{T} \; + \; 
 \frac{1}{2} \sum_{a,b=1}^{N_{2}} L_{a} L_{b} \hat{T}^{-1}_{ab}
 \Biggr\}
\label{45}
\end{eqnarray}
where
\begin{eqnarray}
 \label{46}
\mbox{Tr} \,\ln \hat{T} &=& N_{2} \ln \gamma \; + \; \ln\Bigl(1 \; + \; N_{2} \frac{\kappa}{\gamma}\Bigr)
 \\
\nonumber
 \\
\hat{T}^{-1}_{ab} &=& \frac{1}{\gamma} \delta_{ab} \; - \; \frac{\kappa}{\gamma\bigl(\gamma + N_{2} \kappa\bigr)}
\end{eqnarray}
Simple calculations yield:
\begin{eqnarray}
\nonumber
&&{\cal Z}_{\epsilon}\bigl( N_{1}, n_{1}, N_{2}, n_{2}, x_{1}, x_{2}, t, \Delta t \bigr) \; = \;
C(N_{1}+N_{2},t) \, C(N_{2}, \Delta t) \; 
\times
 \\
\nonumber
 \\
 \nonumber
&& \hspace{10mm} \times \exp\Biggl\{
 -\frac{1}{2} \beta G
 -\frac{1}{2} N_{2} \ln \gamma  -  \frac{1}{2} \ln\Bigl(1 + N_{2} \frac{\kappa}{\gamma}\Bigr)
 -\frac{L^{2} N_{2}}{2(\gamma + N_{2}\kappa)} 
+ \frac{\beta L}{\gamma\Delta t}\bigl(N_{2} x_{2} + n_{2}\epsilon\bigr) -
\\
\nonumber
 \\
&& \hspace{10mm} - \frac{\beta\kappa N_{2}(N_{2} x_{2} + n_{2}\epsilon)}{
                         \gamma \Delta t (\gamma + N_{2} \kappa)} 
                 - \frac{\beta^{2}\kappa (N_{2} x_{2} + n_{2}\epsilon)^{2}}{
                        2 \gamma (\Delta t)^{2} (\gamma + N_{2} \kappa)}
         + \frac{\beta^{2}}{2\gamma (\Delta t)^{2}} \Bigl[(N_{2}-n_{2})x_{2}^{2} + n_{2}(x_{2}+\epsilon)^{2}\Bigr]
\Biggr\}
\label{48}
\end{eqnarray}

Next step of the calculations is to take the limits $N_{1,2}\to 0$. Using explicit expressions
(\ref{14})-(\ref{16}), (\ref{38}), (\ref{41}) and (\ref{43}), one easily finds:\
\begin{eqnarray}
 \label{49}
 \lim_{N\to 0} C(N,t) &=& 1
 \\
\nonumber
 \\
  \label{50}
 \lim_{N\to 0} A(N,t) &=& - \frac{1}{3}\beta u t^{2}
  \\
\nonumber
 \\
 \label{51}
 \lim_{N\to 0} B(N,t) &=&  \frac{1}{2}\beta u t^{2}
   \\
\nonumber
 \\
 \label{52}
  \lim_{N_{1},N_{2}\to 0} \beta G &=& 
           \frac{\beta n_{1}\epsilon}{t} (2x_{1} + \epsilon)
          +\frac{\beta n_{2}\epsilon}{\Delta t} (2x_{2} + \epsilon)
          -\frac{1}{3} (\beta n_{1}\epsilon)^{2} u t
          -\frac{1}{3} (\beta n_{2}\epsilon)^{2} u \Delta t
 \\
\nonumber
 \\
 \label{53}
 \lim_{N_{1},N_{2}\to 0} L &=& \frac{1}{3} \beta^{2} n_{1} \epsilon u t 
                             + \frac{1}{6} \beta^{2} n_{2} \epsilon u \Delta t
 \\
\nonumber
 \\
 \label{54}
\lim_{N_{1},N_{2}\to 0} \kappa &=& -\frac{1}{3} \beta^{2} u (t + \Delta t) 
\end{eqnarray}
Substituting the above limiting values into eq.(\ref{48}) we get
\begin{eqnarray}
\nonumber
&& \lim_{N_{1,2}\to 0} {\cal Z}_{\epsilon}\bigl( N_{1}, n_{1}, N_{2}, n_{2}, x_{1}, x_{2}, t, \Delta t \bigr)
   \; \equiv \; 
   Z_{\epsilon}\bigl(n_{1}, n_{2}, x_{1}, x_{2}, t, \Delta t \bigr) \; =
 \\
\nonumber
 \\
&&=\exp\Biggl\{ 
          -\frac{\beta n_{1}\epsilon}{2t} (2x_{1} + \epsilon)
          -\frac{\beta n_{2}\epsilon}{2(t + \Delta t)} (2x_{2} + \epsilon)
          +\frac{1}{6} (\beta n_{1}\epsilon)^{2} u t
          +\frac{1}{6} (\beta n_{2}\epsilon)^{2} u (t + \Delta t)
          + \frac{(\beta n_{1}\epsilon)(\beta n_{2} \epsilon) u t^{2}}{3(t + \Delta t)} 
\Biggr\}
\label{55}
\end{eqnarray}
Substituting the above result into eq.(\ref{32}) and introducing notations
$\beta n_{1,2} \epsilon \; = \; s_{1,2}$ in the limit $\epsilon \to 0$ we obtain:
\begin{eqnarray}
 \nonumber
&& \int\int_{-\infty}^{+\infty} d v_{1} d v_{2} \; P_{x_{1},x_{2} t, \Delta t} (v_{1}, v_{2})
 \exp\bigl\{ s_{1} v_{1} + s_{2} v_{2}\bigr\} 
 \; = \;
 \\
\nonumber
 \\
\label{56}
&&=\exp\Biggl\{ 
 -\frac{x_{1}}{t} s_{1} - \frac{x_{2}}{t + \Delta t} s_{2} 
 +\frac{1}{6} u t s_{1}^{2} + \frac{1}{6} u (t + \Delta t) s_{2}^{2}
 +\frac{u t^{2}}{3(t + \Delta t)} s_{1} s_{2}\
\Biggr\}
\end{eqnarray}
Redefining
\begin{eqnarray}
 \label{57}
 s_{1} &=& \sqrt{\frac{3}{ut}} \; \omega_{1}
\\
\nonumber
 \\
\label{58} 
 s_{2} &=& \sqrt{\frac{3}{u(t+\Delta t)}} \; \omega_{2}
\\
\nonumber
 \\
\label{59}  
v_{1} &=& -\frac{x_{1}}{t} \; + \; \sqrt{\frac{1}{3} u t} \; \tilde{v}_{1}
\\
\nonumber
 \\
\label{60}  
v_{2} &=& -\frac{x_{2}}{t+\Delta t} \; + \; \sqrt{\frac{1}{3} u (t+\Delta t)} \; \tilde{v}_{2}
\\
\nonumber
 \\
\label{61}
\Delta t &=& \xi \, t
\end{eqnarray}
we get the following relation for the probability distribution function 
${\cal P}_{\xi}\bigl(\tilde{v}_{1}, \, \tilde{v}_{2}\bigr)$ for the rescaled
velocities $\tilde{v}_{1}$ and $\tilde{v}_{2}$, eqs.(\ref{59})-(\ref{60}):
\begin{equation}
 \label{62}
\int\int_{-\infty}^{+\infty} d\tilde{v}_{1} \tilde{v}_{2} {\cal P}_{\xi}\bigl(\tilde{v}_{1}, \, \tilde{v}_{2}\bigr)
\exp\bigl\{ \omega_{1} \tilde{v}_{1} + \omega_{2} \tilde{v}_{2}\bigr\} 
 \; = \; 
\exp\Bigl\{
\frac{1}{2} \omega_{1}^{2} + \frac{1}{2} \omega_{2}^{2} + \frac{\omega_{1} \omega_{2}}{(1+\xi)^{3/2}}
\Bigr\}
\end{equation}
Performing simple inverse Laplace transformation
\begin{equation}
 \label{63}
{\cal P}_{\xi}\bigl(\tilde{v}_{1}, \, \tilde{v}_{2}\bigr) \; = \;
\int\int_{-i\infty}^{+i\infty} \frac{d\omega_{1} d\omega_{2}}{(2\pi i)^{2}}
\; 
\exp\Bigl\{
\frac{1}{2} \omega_{1}^{2} + \frac{1}{2} \omega_{2}^{2} + \frac{\omega_{1} \omega_{2}}{(1+\xi)^{3/2}}
-\omega_{1} \tilde{v}_{1} - \omega_{2} \tilde{v}_{2}
\Bigr\}
\end{equation}
one eventually obtain the following very simple result for the two-time velocities distribution function:
\begin{equation}
 \label{64}
{\cal P}_{\xi}\bigl(\tilde{v}_{1}, \, \tilde{v}_{2}\bigr) \; = \;
\frac{1}{2\pi} \sqrt{\frac{(1+\xi)^{3}}{(1+\xi)^{3} - 1}} \; 
\exp\Biggl\{
-\frac{(1+\xi)^{3}}{2\bigl[(1+\xi)^{3} - 1\bigr]}
\Bigl(
\tilde{v}_{1}^{2} -2 \frac{\tilde{v}_{1}\tilde{v}_{2}}{(1+\xi)^{3/2}} + \tilde{v}_{2}^{2}
\Bigr)
\Biggr\}
\end{equation}
where $\xi = \Delta t/t$ is the reduced separation time parameter.

One can easily check that in the limit of infinite separation time, $\xi \to \infty$,
the distributions of two velocities are getting independent:
\begin{equation}
 \label{65}
\lim_{\xi\to\infty} {\cal P}_{\xi}\bigl(\tilde{v}_{1}, \, \tilde{v}_{2}\bigr) \; = \;
\frac{1}{2\pi} \; 
\exp\Bigl\{ -\frac{1}{2} \tilde{v}_{1}^{2} - \frac{1}{2} \tilde{v}_{2}^{2}
\Bigr\}
\end{equation}
while in the opposite limit of coinciding times, $\xi \to 0$, one finds
\begin{equation}
 \label{66}
{\cal P}_{0}\bigl(\tilde{v}_{1}, \, \tilde{v}_{2}\bigr) \; = \;
\frac{1}{\sqrt{2\pi}} \; 
\exp\Bigl\{ -\frac{1}{2} \tilde{v}_{1}^{2} \Bigr\} \; 
\delta\bigl(\tilde{v}_{1} - \tilde{v}_{2}\bigr)
\end{equation}
as it should be.

Besides, using the exact result, eq.(\ref{64}) one can easily compute the time 
dependence of the two velocities correlation function:
\begin{equation}
 \label{67}
\langle \tilde{v}_{1} \tilde{v}_{2} \rangle \; = \; (1 + \xi)^{-3/2}
\end{equation}
as well as the probability distribution function for the velocities difference
$\tilde{v} \equiv \tilde{v}_{2} - \tilde{v}_{1}$:
\begin{equation}
 \label{68}
{\cal P}_{\xi}\bigl(\tilde{v}\bigr) \; = \;
\frac{1}{2\pi} \sqrt{\frac{(1+\xi)^{3/2}}{4\pi \bigl[(1+\xi)^{3/2} - 1\bigr]}} \; 
\exp\Biggl\{
-\frac{(1+\xi)^{3/2}}{4\bigl[(1+\xi)^{3/2} - 1\bigr]} \, \tilde{v}^{2}
\Biggr\}
\end{equation}

 \section{Conclusions}

In this paper we have considered the problem of velocity distribution functions 
in the Burgulence problem in terms of the toy Gaussian model of (1+1) directed polymers.
In particular the exact result for the two-time free energy, eq.(\ref{28}),
and two-time velocity distribution functions, eq.(\ref{64}) has been derived. 
Of course the considered system is too far from the realistic one. Nevertheless, it has one important
advantage: being exactly solvable,  some of its statistical properties are 
rather non-trivial. 
All that, in my view, makes this model to be rather useful tool for testing new ideas and various 
technical aspects of the calculations (like the replica technique considered in this paper).

Following the proposed route, the next step  would be to consider the model
with finite range correlations of the random potentials. Of course, one can not hope to
get exact results here.
In terms of the replica approach, first of all, one is facing the problem 
of $N$-particle  quantum bosons with attractive {\it finite range} interactions 
whose solution is not known. Nevertheless even the qualitative understanding of the
structure of the $N$-particle wave function of this system
(which at the qualitative level might be not so much different from that of the Bethe ansatz solution 
for the $\delta$-correlated potentials)
could hopefully be sufficient to get some understanding of the velocity statistics 
in the Burgulence problem.

\acknowledgments

This work was supported in part by the grant IRSES DCPA PhysBio-269139.



\begin{thebibliography}{99}

\bibitem{KPZ} M.Kardar, G.Parisi, Y-C.Zhang,
   Phys.\ Rev.\ Lett.\ {\bf 56}, 889 (1986)



\bibitem{hh_zhang_95} T.\ Halpin-Healy and Y-C.\ Zhang,
   Phys.\ Rep.\ {\bf 254}, 215 (1995).


\bibitem{burgers_74} J.M.\ Burgers, {\it The Nonlinear
   Diffusion Equation} (Reidel, Dordrecht, (1974)).

\bibitem{kardar_book} M.\ Kardar,
   "Statistical physics of fields" (Cambridge: Cambridge University Press, (2007))


\bibitem{hhf_85} D.A.\ Huse, C.L.\ Henley, and D.S.\ Fisher,
    Phys.\ Rev.\ Lett.\ {\bf 55}, 2924 (1985).

\bibitem{numer1} D.A.\ Huse and C.L.\ Henley,
    Phys.\ Rev.\ Lett. {\bf 54}, 2708 (1985).

\bibitem{numer2} M.\ Kardar and Y-C.\ Zhang,
    Phys.\ Rev.\ Lett.\  {\bf 58}, 2087 (1987).


\bibitem{kardar_87} M.\ Kardar,
   Nucl.\ Phys.\ {\bf B 290}, 582 (1987).


\bibitem{bouchaud-orland} J.\ P.\ Bouchaud and H.\ Orland,
   J.\ Stat.\ Phys.\ {\bf 61}, 877 (1990)

\bibitem{Brunet-Derrida} E.\ Brunet and B.\ Derrida,
   Phys. \ Rev. \ E {\bf 61}, 6789 (2000)

\bibitem{Johansson}  K.\ Johansson,
   Comm. \ Math. \ Phys. \ {\bf 209}, 437 (2000)

\bibitem{Prahofer-Spohn} M.\ Prahofer and H.\ Spohn
   J.\ Stat.\ Phys.\ {\bf 108}, 1071 (2002)

\bibitem{Ferrari-Spohn1} P.\ L.\ Ferrari and H.\ Spohn,
   Comm. \ Math. \ Phys. \ {\bf 265}, 1 (2006)


\bibitem{KPZ-TW1a} T.Sasamoto and H.Spohn,
         Phys.\ Rev.\ Lett.\ {\bf 104}, 230602 (2010)

\bibitem{KPZ-TW1b} T.Sasamoto and H.Spohn,
         Nucl.\ Phys.\ {\bf B834}, 523 (2010)

\bibitem{KPZ-TW1c} T.Sasamoto and H.Spohn,
         J.\ Stat.\ Phys. {\bf 140}, 209 (2010)

\bibitem{KPZ-TW2}  G.Amir, I.Corwin and J.Quastel,
        Comm.\ Pure Appl.\ Math.\ {\bf 64}, 466 (2011)

\bibitem{BA-TW1} V.Dotsenko and B.Klumov,
         J.Stat.Mech. P03022 (2010)

\bibitem{BA-TW2} V.Dotsenko,
         EPL, {\bf 90},20003 (2010)

\bibitem{BA-TW3}  V.Dotsenko,
         J.Stat.Mech. P07010 (2010)

\bibitem{LeDoussal1} P.Calabrese, P. Le Doussal and A.Rosso,
         EPL, {\bf 90},20002 (2010);
         
\bibitem{LeDoussal2} P.Calabrese and P. Le Doussal,
         Phys.\ Rev.\ Lett.\ {\bf 106}, 250603 (2011);
         arXiv:1204.2607
         
\bibitem{end-point} V.Dotsenko,
         J. Stat. Mech. P02012 (2012)            

\bibitem{goe} V.Dotsenko,
         J. Stat. Mech. P11014 (2012)
         
\bibitem{LeDoussal3} T.Gueudr\'e and P. Le Doussal,
        EPL, {\bf 100}, 26006 (2012).
         
        
 \bibitem{Corwin} I.\ Corwin,
   "The Kardar-Parisi-Zhang equation and the universality class",
   arXiv:1106.1338, Random Matrices: Theory Appl. {\bf 1}, 1130001 (2012)


\bibitem{Borodin} A.Borodin, I.Corwin and P.Ferrari,
       {\it Free energy fluctuations for directed polymers in random media in 1+1 dimension},
        arXiv:1204.1024 (2012)       



\bibitem{Prolhac-Spohn} S.\ Prolhac and H.\ Spohn,
         J.Stat.Mech. P01031 (2011)
         
\bibitem{2pointPDF} V.Dotsenko, 
         J.Phys. A: Math. Theor. {bf 46}, 355001 (2013)         

\bibitem{Imamura-Sasamoto-Spohn} T.Imamura, T.Sasamoto and H.Spohn,
         J.Phys. A: Math. Theor. {bf 46}, 355002 (2013)


\bibitem{Prolhac-Spohn-N} S.\ Prolhac and H.\ Spohn,
         J.Stat.Mech. P03020 (2011)

\bibitem{2time} V.Dotsenko, 
         J.Stat.Mech. P06017 (2013)\
         
\bibitem{Npoint} V.Dotsenko, 
         Cond.Mat.Phys. {\bf 17}, 33003 (2014)    
         

\bibitem{Sinai} Ya. G. Sinai, 
         Commun.Math.Phys. {\bf 148}, 601 (1992); 
         J.Stat.Phys. {\bf 64}, 1 (1991)
         
\bibitem{Bouch-Mez-Par} J.P.Bouchaud, M.Mezard and G.Parisi,
         Phys.Rev. E {\bf 52}, 3656 (1995)
         
\bibitem{burgulence} J.Bec and K.Khanin, 
         Physics Reports {\bf 447}, 1 (2007)        

\bibitem{Larkin} A. I. Larkin, 
         Sov. Phys. JETP {\bf 31}, 784 (1970)
 
\bibitem{Larkin-Ovchinnikov} A. I. Larkin anf Yu. N. Ovchinnikov,
         J. Low Temp. Phys. {\bf 34}, 409 (1979)
 
\bibitem{Gorochov-Blatter} D. A. Gorokhov and G. Blatter,
         Phys. Rev. Lett. {\bf 82}, 2705 (1999)


         
\bibitem{gaussian}  V.S.Dotsenko, V.B.Geshkenbein, D.A.Gorokhov and G.Blatter, 
         Phys.Rev. {\bf B82}, 174201 (2010)
        
         
\bibitem{replicas} V.Dotsenko,
         Phylosophical Magazine, {\bf 92}, 16 (2012)




\end{thebibliography}
\end{document}